Title:
*Building a Framework for Indigenous Astronomy Collaboration, "Native Skywatchers, Indigenous Scientific Knowledge Systems, and The Bell Museum"*

Annette S. Lee, Sally Brummel, Kaitlin Ehret, Sarah Komperud, Thaddeus LaCoursiere
July 2020

Lee, Annette S aslee@stcloudstate.edu
Sally Brummel sbrummel@umn.edu
Kaitlin Ehret ehret017@umn.edu
Sarah Komperud komp0030@umn.edu
Thaddeus LaCoursiere tlacours@umn.edu

---

Land Acknowledgement – *The Bell Museum sits on the traditional and treaty land of the Dakota people who, along with the Ojibwe people, are the Indigenous peoples of this land, Mnisóta Makhóčhe, or Minnesota.*

## 1  Introduction – Background to the Project

Hundreds of years ago, colonization happened. Today we are still living out the ripple effects of this history, called by some 'historical trauma' (Gone 2009; Brown-Rice 2013). How does this relate to science, informal science education, and institutions that promote science communication? What obligations or considerations should a science museum have before integrating Indigenous knowledge into their existing programming? Presented in this document is the process of building a framework intended to provide a roadmap for developing Indigenous astronomy programming which can be a model for other institutions that may be interested in collaborating with Indigenous communities. It was written by members of a collaborative project between two stakeholders: an Indigenous astronomer-scholar and Minnesota's official natural history museum and planetarium

Here in *Mnisóta Makhóčhe*, where the 'Water Reflects the Sky', or Minnesota (Westerman and White 2012), there are four Dakota reservations and seven Ojibwe reservations (Peacock and Day 2000).  The Dakota tribes here in Minnesota are:  Lower Sioux Indian Community, Prairie Island Indian Community, Shakopee Mdewakanton Sioux (Dakota) Community, and Upper Sioux Community. The Ojibwe tribes are: Bois Forte Band of Chippewa, Fond du Lac Reservation, Grand Portage Band of Chippewa Indians, Leech Lake Band of Ojibwe, Mille Lacs Band of Ojibwe, Red Lake Band of Chippewa Indians, White Earth Reservation (mn.gov n.d.). Note the less respectful, anglicized word 'Sioux' has been used to refer to the Dakota people. The etymology of the word 'Sioux' is from a corrupted French interpretation, Nadouessioux, of the Ojibwe word Naudoway, 'Enemy' (Peacock and Day 2000). Similarly, the less respectful, anglicized word 'Chippewa' has been used to refer to the Ojibwe people. Both 'Sioux' and 'Chippewa' are sometimes still in use because treaties and other historical documents were originally written using these terms. Lastly, the more general term, Anishinaabe, or "the People" is sometimes used by Ojibwe communities (UMN 2020).

Historically the Dakota lived in the area and later (about 1500 years ago) the Ojibwe migrated into Minnesota from the east coast due to warfare and prophecy. The prophecy told of a 'land where the food grows on the water', i.e. *manoomin,* wild rice (Peacock and Day 2000; MNHS n.d.; n.d.). The Dakota people of the land that is now known as Minnesota have four groups: *Bdewakantunwan* (Mdewakanton), *Wahpetunwan* (Wahpeton), *Wahpekute*, and *Sissitunwan* (Sisseton). Two other related groups are Lakota and Nakota, (Teton, Yanktonai, and Yankton), which are located in South Dakota, North Dakota, Nebraska, and Canada. All together there are seven groups, called the '*Oceti Ŝakowiŋ*' or the Seven Council Fires (MNHS n.d.). In 1826 there was a treaty between the Ojibwe and Dakota negotiated by the U.S. government that effectively protected the fur trade industry by creating the boundary between Ojibwe and Dakota. The Dakota were pushed to the southwest of Minnesota, where woodlands changed to prairie.

Nearly two hundred years later, there are eleven reservations in the state of Minnesota which represent only a small fraction of the traditional homelands. Today there is a population of approximately 61,000 American Indians or Alaska native in the state or 1.1% of the population, although the largest multiracial population in the U.S. is mixed Native American and White (cultureconnection.org n.d.). Note that the terminology varies. 'Native American' is widely used in the U.S., although in some cases the outdated term 'Indian' is used within a legal context. The term 'Indigenous' is a broader term used to include a variety of tribes or groups and has a connotation of a global or international context. Keep in mind that by some estimates, over a two hundred year period, up to 95% of the Indigenous population was annihilated: "In 1491, about 145 million people lived in the Western Hemisphere. By 1691, the population of Indigenous Americans had declined by 90-95 percent, or by around 130 million people" (McKenna and Pratt 2014, 375). Despite this catastrophic loss there are thriving Native communities where language and culture have deep meaning and value. Revitalization efforts are strong.

## 2  What is the problem – What is our solution

Outside of the Native communities, the language revitalization efforts are not well known, and barely understood are colonization and the need for revitalization. A museum can provide a platform for Indigenous people to introduce and discuss these topics with the museum visitors, both Indigenous and non-Indigenous.

The Bell Museum was established by the state legislature in 1872 and is held in trust by the University of Minnesota to preserve and interpret the state's rich natural history. In summer 2018 the Bell Museum opened a new $79.2 million facility on the University of Minnesota's St. Paul campus. This public-private investment in Minnesota's natural history museum was predicated on the idea of public education through authentic and immersive experiences that celebrate Minnesota's natural heritage and emphasize both the process and outcomes of science.

The museum hosts the state's largest fully digital public planetarium, offers a far-reaching portable planetarium program that travels throughout Greater Minnesota, hosts monthly star parties on the observation deck, and provides public access to the latest astronomy research at the University of Minnesota.

The Bell Museum has worked to support Indigenous revitalization efforts by providing a platform for Native interpretation of perspectives including traditional ecological knowledge in its galleries, Indigenous storytelling events, and features on Native American science researchers and artists throughout the museum. These efforts have been conducted in collaboration with Native people with the goal of honoring the significant and continued expertise of Minnesota's Indigenous peoples, focusing on both traditional ecological knowledge and contributions to creating more sustainable futures.

It was a welcome surprise that these efforts have been extremely well-received by the museum's non-Indigenous community, with enthusiastic requests for more opportunities to learn from Native Americans and about Native ecological and astronomy knowledge. The Bell Museum planetarium staff heard from people who wanted the opportunity to learn about and interact with programming that is rooted in the Indigenous, especially Ojibwe and Dakota perspective. Yet, a clear path to collaboration was not immediately obvious. Bell Museum planetarium staff desired to develop a framework to guide their process. The questions they had, and continue to explore in approaching this framework are: what is this systematic approach for developing Indigenous astronomy programming, and can they, as non-Indigenous people, present the programming, or should it be in concert with members of the community?

**2.1 Relationship building**

The idea for collaboration with *Native Skywatchers* was rooted in a motivation for authentic collaboration with Indigenous astronomy and the Bell Museum. Over a decade ago, Annette Lee founded the *Native Skywatchers* (Lee 2007) research and programming initiative, a grassroots Indigenous lead effort to revitalize the star knowledge of the Ojibwe and D/Lakota peoples. Lee is also an associate professor of astronomy and planetarium director at St. Cloud State University. Her communities are Ojibwe and D/Lakota and she is mixed-race Lakota, family name *Wanbli Luta* (Red Eagle).

It is in these first exploratory conversations that the seeds of a relationship were nurtured. When A. Lee was heard speaking on the NPR Science Friday broadcast, "Relearning the Star Stories of Indigenous People: How the lost constellations of indigenous North Americans can connect culture, science, and inspire the next generation of scientists." (Taylor 2019), it inspired a conversation between the Indigenous astronomer, A. Lee, and the executive director at the Bell Museum. Without a doubt, building a healthy, vibrant relationship with Indigenous partners can take time, but there is no substitute. Leroy Little Bear, Blackfoot elder and scholar, explains the importance of relationships in Indigenous culture, "It's about all my relations. Everything is inter-related…..Blackfoot science is largely based on relations. You can say, If it is not about relations, (then) it is not (Indigenous) science" (Little Bear 2020). The first and most important step in bringing in Indigenous science into the museum/planetarium setting is to build a relationship with the Indigenous community. Historically the 'grab and go' model allowed academia, including museums, to address Indigenous communities and their cultural artifacts as objects. This is in direct opposition to the model described here. Indigenous culture comes with a living voice and interpretation. It is imperative to seek out Indigenous knowledge keepers and work to build healthy relationships with them. This includes allowing the integration of the Indigenous science to be led by Indigenous people, the living voice is what matters here.

Another aspect of 'relationship building' is the relationship with the land. In recent times more institutions and individuals are recognizing the need to acknowledge the Indigenous people of the land that the institution exists on. If the information of the local tribe in your area is not obvious then it can be researched using various websites ("NativeLand.Ca" 2020). The Bell Museum has a welcome panel in the lobby of the museum, made by a Dakota person to Dakota land, and incorporates a land acknowledgement practice for public programs.

Lastly, the time factor needed to build an authentic relationship with Indigenous communities is an important consideration. Relationships that are genuine and sustainable are not quick and simple, but rather usually take years to develop.

Learning about the importance of relationships in Indigenous culture with people and with the land, and understanding the time needed to work on building these relationships, provide a foundation for our work to develop astronomy programming with Indigenous partners.

**2.2 Indigenous Science**

Often times planetarium professionals, although highly credentialed in the fields of science and education, are not necessarily aware of the overlying history of colonization, race, and culture as it applies to Indigenous science. As museums begin to rethink and re-examine their content, exhibits, artifacts, and programming a wider lens must be considered.

The first step in creating a more inclusive, diverse, and equitable learning environment in planetariums, and indeed in all of STEM, is to acknowledge that science is by definition embedded with culture and history. Note that 'culture' is more than superficial or performative markers like holidays, customs, and ethnic food. Culture is often unconscious but has a huge impact on the philosophical underpinnings of a society (Lee 2020). Similarly, science has only become the ever narrowing, laser focused, divide-and-conquer endeavor that has evolved over the past few hundred years to what we know today. For example, the etymology of the word 'physics' comes from 'physica' or 'physicks' which is from Latin meaning 'the philosophy of nature or natural philosophy' (Oxford English Dictionary 2019). Surely we can recognize that the multitude of peoples that have existed on Earth have had different ways of 'relating to nature'. And that by acknowledging one and only one culture's natural philosophy we are losing other valuable perspectives. Indeed, we can recognize that Indigenous people had and still have a keen awareness of the natural world, and go further to point out that relationships and participation with the natural world are key elements of Indigenous science.

The terms "Indigenous science" and "Indigenous Scientific Knowledge Systems" come from the broader framework of "Indigenous Knowledge Systems (IKS)" that is rooted in post-apartheid South Africa, or the 'African Renaissance'.

> *The word, "indigenous" refers to the root of things, and hence to something natural and innate to a specific context, for example, an indigenous plant or tree which grows naturally in the existing environment… IKS in general refers to intricate knowledge systems acquired over generations by communities as they interact with their environment.* (Higgs and van Niekerk 2002, 38)

Similar movements followed in Australia, New Zealand, and more recently Canada. The South African government began to recognize IKS as a critical and valuable perspective worth protecting and promoting at the federal level. Indigenous African scholars like Munyaradzi Mawere, Lesley LeGrange, and Meshach Ogunniyi (Le Grange 2007; Mawere 2015; Ogunniyi 2005) give the following outline of the differences between Indigenous Knowledge Systems and Western Science:

*Table 1 – Comparison of Indigenous Knowledge Systems with Western Science, Le Grange 2007, 585*

| Indigenous Knowledge Systems: | Western Science: |
| --- | --- |
| nature is real and partly observable | versus nature is real and observable |
| events have both natural and unnatural causes | versus all events have natural causes |
| the universe is partly predictable and partly unpredictable | versus the universe is predictable |
| language is important as a creative force in both the natural and unnatural worlds | versus language is not important to the workings of the natural world |
| knowledge is a critical part of culture | versus science is culture free |
| humans are capable of understanding only part of nature | versus humans are capable of understanding nature |

To be clear, Indigenous Astronomy is a branch of the more general term, Indigenous Knowledge Systems (IKS). Here are some additional elements of Indigenous science by Blackfoot elder and scholar, LeRoy Little Bear:

(1) all things are made of energy, "In Aboriginal philosophy, existence consists of energy. All things are animate, imbued with spirit, and in constant motion" (Little Bear 2000, 77); (2) we are all related, "In this realm of energy and spirit, interrelationships between all entities are of paramount importance…" (Little Bear 2000, 77).

This is in contrast with the Western science worldview: humans exercise dominion over nature to use it for personal and economic gain; human reason transcends the natural world and can produce insights independently; nature is completely decipherable to the rational human mind (Knudtson and Suzuki 1992). As (Carter et al. 2003) note, "The emphasis of Western science is on mastering, controlling, and transforming nature and promotes individual success and competition" (p. 6).

One of the leading voices of Native science, Gregory Cajete states, "Native science is born of a lived and storied participation with the natural landscape. To gain a sense of Native science one must participate with the natural world" (Cajete 2000, 2). Two of the fundamental views of science using the Native cultural lens are: relationship and participation.

So how does this relate to museums? The first reason museums should embrace Indigenous science is interest. These museum-goers have expressed interest in t programming that relates to the Indigenous people of the place, namely Dakota and Ojibwe people. If the programming, exhibits, and planetarium shows, at a state-of-the-art planetarium and natural history museum are not relevant, then people will not show up. The second reason museums should include Indigenous science is because it is equitable, inclusive, and ethical. Indeed, the big picture demands it. In light of the COVID-19 global pandemic, huge economic pressures on museums, and the Black Lives Matter social justice movement, there is a growing awareness that museums to some extent have upheld colonial histories and values in their statues, collections, and academic pursuits. This is no longer acceptable, and museums are in need of a social justice update that continues beyond repatriation (CNN 2020).

## 3  Objectives of the Project  & Specific Actions

### 3.1  Goals

Early on, it was decided that the collaboration should have very clear goals. The following are the Core Functions and Goals of this project:

- Participate in the development of a framework (and best practices) for Bell Museum astronomy educators to share Indigenous knowledge in respectful and authentic ways throughout its suite of astronomy efforts. One of the most significant questions to be addressed is how non-Native astronomy educators can most appropriately promote and educate about Indigenous star knowledge. We expect this consultancy to answer this question specific to the Bell Museum and create a roadmap that will guide future efforts.
- Design and deliver foundational educational training of Bell Museum astronomy educators about Indigenous star knowledge. Annette Lee and the *Native Skywatchers* program she founded is an educational program designed for professional educators and others, and we would expect the Bell astronomy educators to benefit from Annette's knowledge and experience.
- The development of assets, such as recorded stories and constellation visuals, that afford Bell Museum astronomy educators to meaningfully integrate Indigenous star knowledge into planetarium shows, star maps, digital education programs, Star Parties (nighttime sky observation events), blog posts, and other means of engaging the public. The framework will guide the most appropriate assets to be built.

- Share Indigenous star knowledge with the public through live online programs (delivered individually or co-presented) throughout the project period. These programs will provide inspiration for the Bell team, as well as opportunities for Bell team to implement and practice the knowledge and frameworks they develop under Annette's supervision.

## 3.2 Deliverables

The following were the tangible deliverables for this collaborative project:

- Weekly Training meetings which served as a regular structure to keep projects moving forward. These meetings served multiple purposes: conversational trainings, provide focus and a forum for project discussions.
- Star Map Booklet. Added selected Indigenous Astronomy content to the existing Star Map Booklet, a bi-monthly publication. This Bell Museum publication highlights astronomical treasures including an 8.5 x 11" star map, Moon phase calendar, and astronomical deep sky objects. This project integrated Indigenous content in two issues: July-Aug and Sept-Oct 2020.
- 'Live' show participation. The show will be a virtual format, geared toward a family audience, completely virtual. The Indigenous consultant will join one or more Bell Planetarium Staff, live stream a thirty-minute show that will be recorded.
- Framework. A written document of this collaboration and best practices published with the intent of sharing some highlights of a unique collaboration offered as guidance and useful for further discussion.

This project had a timeline of three summer months. The reality of the COVID-19 pandemic required meetings and work to be done remotely. The team is on track to meet all the deliverables. Assessment of this work was primarily summative, although some anecdotal evidence of engagement was formative.

## 4 Q&A and Next Steps

This section presents questions and concerns that Bell Planetarium staff had at the beginning of the project. Specific questions and answers are shared here for dissemination and learning. In general, the best strategy is to ask the Indigenous partner/s directly what their perspectives are on any of the questions. Expect a spectrum of answers. Realize that the Indigenous voice should be the lead voice on the integrating of Indigenous content into the museum/planetarium setting. The process should be an ongoing discussion.

**Question:** *What is the acceptable name for Indigenous People? Native American vs. First Nation vs. Indian vs. American Indian… Which term is more appropriate, more respectful?*
**Answer:** Generally speaking in the U.S. the term Native American is used widely, but there are some that prefer American Indian. In Canada the preferred term is 'First Nations' and to be accurate it should be stated 'First Nations and Metis' (Kesler, Crey, and Hanson 2009). 'Indigenous' is often used, particularly on a global scale and seems to be gaining in popularity.

**Question:** *What motivates this project? Why do this project?*
**Answer:** From the Indigenous Consultant's (A. Lee) point of view, the project motivation is rooted in the need for resource-rich, regional museums (like the Bell Museum) to practice cultural inclusion and relevancy. It makes sense demographically and it is the honorable thing to do.

**Question:** *Why hasn't a project like this been done before?*
**Answer:** Due to the history of colonization, there has been centuries of loss of knowledge, language, and culture. In addition, academic institutions have only recently begun to recognize Indigenous Knowledge Systems (Higgs and van Niekerk 2002).

**Question:** *What is the bigger picture? Why is culture relevant?*
**Answer:** 'Culture' is more than superficial or performative markers like holidays, customs, and ethnic food. Culture is often unconscious but has a huge impact on the philosophical underpinnings of a society (Lee 2020). As long as science is done by human beings, science is embedded with culture. Acknowledging culture is like acknowledging the air we need to breathe. Once in a while we can see our breath, (like on a frosty winter day in *Mnisóta Makhóčhe* - Minnesota), but it is always present. As long as we are human beings, culture is our humanity. Everyone can agree that being human is relevant to everything we do.

**Question:** *What are some challenges in communicating the Indigenous knowledge in a planetarium setting?*
**Answer:** Many tribes have lost part or all of this knowledge. At the same time it was considered sacred knowledge. Traditionally there was/is protocol around what is being shared, by who, to whom, and when. Planetarium educators should practice humility and compassion, recognizing this loss of cultural knowledge as a trauma (Walkley and Cox 2013). Conversely, some non-Indigenous visitors might not understand why the Indigenous knowledge is included in a 'mainstream' planetarium show. Here is a great opportunity to get an important dialog going on cultural inclusion and diversity.

At the time of this submission, the collaboration is still in progress. One next step is to use this paper as a basis for the framework, which discusses the best practices, key questions and relationships, approaches, and protocols for incorporating Indigenous star knowledge into their programs and content, and as a model that another institution who might be interested in this process can follow.

**5 Conclusion**

Lastly, there are two important concluding points: (1) protocol, and (2) two-eyed seeing. First, protocol means that there is an established system of rules and restrictions around cultural and tribal knowledge that needs to be acknowledged and respected in order for the content to be accurate and authentic. Protocol varies between tribes and subgroups. Follow the guidelines of the Indigenous consultant or collaborator. Acknowledge the source of the teaching. The idea of protocol is addressed in the D/Lakota Constellation Guidebook:

> *We encourage mindfulness of cultural protocols. Native knowledge is sometimes a different way of knowing than Western science. There are strict cultural protocols that must be respected, such as when some stories are to be told; for example, some are only told when there is snow on the ground. We must be extremely careful not to introduce or propagate error into the written or oral records. Use caution and be hesitant. Users of these materials are urged to seek out elders and native community members to bring into the classroom. Materials represented here should be viewed as a beginning.* (Lee, Rock, and O'Rourke 2014)

Traditionally, Indigenous astronomy was (and still is) considered a sacred knowledge where it was sometimes protected or cared for by certain people. For example in Lakota, traditionally families with the name '*Luta*' or Red were responsible for the upholding the star knowledge (Lee et al. 2012). Unique people were designated to be involved with the knowledge at the deepest level. At the same time, everyone was a part of the knowledge available in the land and in the sky.

Secondly, an important guiding principle in Indigenous Astronomy revitalization efforts is called, '*Etuaptmumk*' or Two-eyed seeing. In the words of two Mi'kmaw elders:

> *Two-Eyed Seeing is learning to see from one eye with the strengths of Indigenous knowledges and ways of knowing, and from the other eye with the strengths of Western knowledges and ways of knowing, and to use both these eyes for the benefit of all.* (Bartlett, Marshall, and Marshall 2012, 336)

This aim of creating collaboration under the framework of '*Etuaptmumk*' or Two-eyed seeing is at least twice as difficult, but is critically important.

Finally, this document serves to support collaborative efforts by museums and planetariums that desire to integrate Indigenous astronomy and science content "in a good way" into their programming, content, and institution. The roadmap starts with tangible efforts to build authentic relationships with Indigenous knowledge keepers and Indigenous scholars. More than a side-note, the Indigenous voice should be allowed to work collaboratively with museum/planetarium staff to lead the integration efforts. Ideally, museums would aim to work away from the 'add on consultant' model towards the more sustainable model of hiring full time positions such as 'Curator of Indigenous Scientific Knowledge Systems' or various other staff positions. The aim here is to increase STEM/Informal Science Learning opportunities for Indigenous youth, adults, and communities, with the hope of increased cultural pride, engagement in science, and community wellness. For the non-Native audience, there is great value in learning and practicing cultural agility. Standing together we have enormous reach and capacity.

# 5 References

Bartlett, Cheryl, Murdena Marshall, and Albert Marshall. 2012. "Two-Eyed Seeing and Other Lessons Learned within a Co-Learning Journey of Bringing Together Indigenous and Mainstream Knowledges and Ways of Knowing." *Journal of Environmental Studies and Sciences* 2 (4): 331–40. https://doi.org/10.1007/s13412-012-0086-8.

Bell Museum. 2020a. "Bell Museum Reopening Q&A | Bell Museum." 2020. https://www.bellmuseum.umn.edu/event/bell-museum-reopening-qa/.

———. 2020b. "Our Response to COVID-19 | Bell Museum." 2020. https://www.bellmuseum.umn.edu/covid19/.

Brown-Rice, Kathleen. 2013. "Examining the Theory of Historical Trauma Among Native Americans." *Professional Counselor* 3 (3).

Cajete, Gregory A. 2000. *Native Science: Natural Laws of Interdependence*. Clear Light Publisher: Santa Fe, New Mexico.

Carter, Norvella P., Patricia J. Larke, Gail Singleton-Taylor, and Erich Santos. 2003. "CHAPTER 1: Multicultural Science Education: Moving Beyond Tradition." *Counterpoints* 120: 1–19.

CNN, Brian Boucher. 2020. "People Are Calling for Museums to Be Abolished. Can Whitewashed American History Be Rewritten?" CNN. June 2020. https://www.cnn.com/style/article/natural-history-museum-whitewashing-monuments-statues-trnd/index.html.

cultureconnection.org. n.d. "American Indians in Minnesota." AMERICAN INDIANS IN MINNESOTA. Accessed June 29, 2020. http://www.culturecareconnection.org/matters/diversity/americanindian.html.

Dilenschneider, Colleen. 2020. "People Intend to Visit Again. Where Will They Go And What Will Make Them Feel Safe? (DATA)." https://www.colleendilen.com/2020/06/10/people-intend-to-visit-again-where-will-they-go-and-what-will-make-them-feel-safe-data/.

Gone, Joseph P. 2009. "A Community-Based Treatment for Native American Historical Trauma: Prospects for Evidence-Based Practice." *Journal of Consulting and Clinical Psychology*, 751–762.

Higgs, P, and M P van Niekerk. 2002. "The Programme for Indigenous Knowledge Systems (IKS) and Higher Educational Discourse in South Africa: A Critical Reflection." *South African Journal of Higher Education*, February 2002 South African Journal of Higher Education 16(3), 16 (3): 12.

Kesler, Linc, Karrmen Crey, and Erin Hanson. 2009. "Terminology." Education website. First Nations and Indigenous Studies- The University of British Columbia. 2009. https://indigenousfoundations.arts.ubc.ca/terminology/.

Knudtson, P, and D. Suzuki. 1992. *Wisdom Of The Elders: Honoring Sacred Native Visions Of Nature*. Toronto: Stoddart Publishing Ltd.

Le Grange, Lesley. 2007. "Integrating Western and Indigenous Knowledge Systems: The Basis for Effective Science Education in South Africa?" *International Review of Education / Internationale Zeitschrift Für Erziehungswissenschaft / Revue Internationale de l'Education* 53 (5/6): 577–91.

Lee, Annette S. 2007. "Native Skywatchers: The Astronomer and Artist." In *Bulletin of the American Astronomical Society*, 39:187. http://adsabs.harvard.edu/abs/2007AAS...210.8001L.

———. 2020. "The Effects on Student Knowledge and Engagement When Using a Culturally Responsive Framework to Teach ASTR 101." Dissertation - research, Cape Town, South Africa: The University of the Western Cape. http://etd.uwc.ac.za/xmlui/handle/11394/7274.

Lee, Annette S., Jim Rock, William Wilson, and Carl Gawboy. 2012. "Red Day Star, the Women's Star and Venus: D (L/N)Akota, Ojibwe and Other Indigenous Star Knowledge." *Science in Society*, 153.

Little Bear, Leroy. 2000. "Jagged Worldviews Colliding." *Reclaiming Indigenous Voice and Vision*, Vancouver, , 77–85.

———. 2020. *Rethinking Our Science: Blackfoot Metaphysics Waiting in the Wings-Reflections by a Blackfoot*. Multimedia | Indigenous Education Institute. virtual. http://indigenouseducation.org/multimedia/.

Mawere, Munyaradzi. 2015. "Indigenous Knowledge and Public Education in Sub-Saharan Africa." *Africa Spectrum* 50 (2): 57–71.

McKenna, Erin, and Scott Pratt. 2014. *American Philosophy-From Wounded Knee to the Present*. New York and London: Bloomsbury Publishing.

mn.gov. n.d. "Minnesota Indian Tribes." Minnesota Indian Tribes. ,. Accessed June 24, 2020. https://mn.gov/portal/government/tribal/mn-indian-tribes/.

MNHS. n.d. "The Dakota People." The Dakota People. Accessed June 24, 2020a. https://www.mnhs.org/fortsnelling/learn/native-americans/dakota-people.

———. n.d. "The Ojibwe People." The Ojibwe People. Accessed June 24, 2020b. https://www.mnhs.org/fortsnelling/learn/native-americans/ojibwe-people.

"NativeLand.Ca." 2020. Native-Land.ca - Our Home on Native Land. 2020. https://native-land.ca/.

Ogunniyi, Mb. 2005. "The Challenge of Preparing and Equipping Science Teachers in Higher Education to Integrate Scientific and Indigenous Knowledge Systems for Learners." *South African Journal of Higher Education* 18 (3): 289–304. https://doi.org/10.4314/sajhe.v18i3.25498.

Oxford English Dictionary. 2019. "Physics, n." In *OED Online*. Oxford University Press. http://www.oed.com/view/Entry/143140.

Peacock, Thomas D., and Donald R. Day. 2000. "Nations within a Nation: The Dakota and Ojibwe of Minnesota." *Daedalus* 129 (3): 137–59.

UMN. 2020. "The Ojibwe People's Dictionary." July 2020. https://ojibwe.lib.umn.edu/search?utf8=%E2%9C%93&q=Anishinaabe&commit=Search&type=ojibwe.

Walkley, M., and T. L. Cox. 2013. "Building Trauma-Informed Schools and Communities." *Children & Schools* 35 (2): 123–26. https://doi.org/10.1093/cs/cdt007.

Westerman, Gwen, and Bruce White. 2012. *Mni Sota Makoce:  The Land of the Dakota*. Minnesota Historical Society Press.